\documentclass{article}
\usepackage{slashed,feynmf,epsfig,amsmath,amssymb,enumitem}
\usepackage[papersize={8.5in,11in}]{geometry}
\begin{document}
\title{Has A 125 GeV Pseudoscalar Resonance Been Observed at the LHC?}
\author{J. W. Moffat\\~\\
Perimeter Institute for Theoretical Physics, Waterloo, Ontario N2L 2Y5, Canada\\
and\\
Department of Physics and Astronomy, University of Waterloo, Waterloo,\\
Ontario N2L 3G1, Canada}
\maketitle
\begin{abstract}
We conjecture that a $125$ GeV resonance $\zeta^0$ with $J^{PC}=0^{-+}$ composed of a quark-antiquark state may have been observed at the LHC. The decay modes of this pseudoscalar particle are determined using the non-relativistic quark model. The leading order partial decay widths for the decays of the pseudoscalar quarkonium resonance $\zeta^0\rightarrow \gamma\gamma$ and $\zeta^0\rightarrow gg$ are compared to the standard model Higgs particle decays $H^0\rightarrow \gamma\gamma$ and $H^0\rightarrow gg$. An experimental analysis at the LHC of the observed production cross sections for the $\zeta^0$ and the Higgs particle could distinguish between a heavy quarkonium $\zeta^0$ particle and a light Higgs particle at $125$ GeV.
\end{abstract}

\begin{fmffile}{ewunifigs}

The excess of events in the $\gamma\gamma$ invariant mass at about $125$ Gev detected in the ATLAS and CMS experiments at the LHC has been tentatively identified as the Higgs particle~\cite{ATLAS1,CMS1,ATLAS2}. It is possible that if the observed $\gamma\gamma$ excess of events becomes statistically significant with increasing data, then this could signal the existence of a pseudoscalar quarkonium resonance bump. A pseudoscalar resonance at around $125$ GeV could be interpreted as a higher energy iteration of the observed lower energy quarkonium meson states~\cite{pdg}. If this proves to be correct, then it may be that a light Higgs particle does not exist and the standard Weinberg-Salam electroweak (EW) model~\cite{Weinberg,Salam} must be revised and a new way of understanding electroweak (EW) symmetry breaking discovered. In recent papers ~\cite{Moffat,Moffat2,Moffat3,Moffat4}, new EW models have been proposed which can be tested at the LHC.

Holdom has recently proposed that condensates of colored fermions drive EW symmetry breaking~\cite{Holdom}. The scenario could give rise to a light pseudoscalar particle. The fermion condensates would require the existence of strong gauge interactions involving either technifermions or a fourth generation of quarks. The pseudoscalar meson would be the lightest state of a new strong interaction sector. It would correspond to a neutral color singlet that arises as a pseudo-Nambu-Goldstone boson, generated by the symmetry breaking. Unlike the Higgs particle, in Holdom's model the pseudoscalar particle does not possess tree level couplings to the $WW$ and $ZZ$, so its branching ratios are significantly suppressed. Other possibilities that have been explored are a fermiophobic Higgs particle~\cite{Gabrielli} for which the fermionic decays such as $b\bar b$ and lepton decays are suppressed, and a vectophobic Higgs particle for which $WW$ and $ZZ$ decays are suppressed~\cite{Gerard}.

The bottomnium and unobserved toponium are isoscalar states $\vert B\rangle=\vert b{\bar b}\rangle$ and $\vert T\rangle=\vert t{\bar t}\rangle$ of heavy quarkonium. With respect to an effective interaction Hamiltonian, heavy quarkonium can appear in two different isosinglet states $\vert\zeta^0\rangle$ and $\vert\zeta^{0'}\rangle$. The effective Hamiltonian is given by ${\cal H}_{\rm eff}={\cal H}_0+{\cal H}_{\rm mass}$, where ${\cal H}_{\rm mass}=K^T{\cal M}K$ and ${\cal M}$ is the mass matrix:
\begin{equation}
\label{Massmatrix}
{\cal M}=\biggl(\begin{array}{cc}m_{\zeta'}&m_{\zeta\zeta'}\\
m_{\zeta\zeta'}&m_{\zeta}\end{array}\biggr).
\end{equation}
Here, $\vert\zeta\rangle$ and $\vert\zeta'\rangle$ are states of quarkonium that interact through the mixing contributions $m_{\zeta\zeta'}$ and $K=\biggl(\begin{array}{cc}\zeta'\\\zeta\\\end{array}\biggr)$.
After diagonalizing the mass matrix and solving for ${\cal M}$, we get the mass formulas
\begin{equation}
m_\zeta=\cos^2\phi m_B+\sin^2\phi m_T,
\end{equation}
and
\begin{equation}
\label{zetaprimemass}
m_{\zeta'}=\cos^2\phi m_T+\sin^2\phi m_B.
\end{equation}
The off-diagonal term is given by
\begin{equation}
m_{\zeta\zeta'}=\cos\phi\sin\phi(m_B-m_T).
\end{equation}
Here, $m_T\sim 2m_t\sim 346$ GeV and $m_B\sim 2m_b\sim 9$ GeV where we have used the measured quark masses: $m_t\sim 173$ GeV and $m_b\sim 4.5$ GeV.

The mixing angle $\phi$ is determined by the equation:
\begin{equation}
\label{mixingangle}
\phi=\arccos[(m_T-m_{\zeta})/(m_T-m_B)]^{1/2}.
\end{equation}
For the mixing angle $\phi\sim 36\,^{\circ}$, we obtain the masses of the quarkonium states $\vert\zeta\rangle$ and $\vert\zeta'\rangle$: $m_{\zeta^0}\sim 125$ GeV and $m_{\zeta^{0'}}\sim 230$ GeV. We identify the new boson resonance discovered at the LHC with the $\zeta^0$ bound state quarkonium resonance. The quarks and anti-quarks are bound together by the QCD gluon force with a corresponding binding energy $E_B$ and the QCD coupling constant $\alpha_s(M_Z)=0.118$~\cite{pdg}.

The short life-time of the top quark for the decay $t\rightarrow bW^+$, $\tau_t\sim 5\times 10^{-25}$ sec., results in the toponium not forming a bound state. However, the toponium state exists as a state of heavy quarkonium.

The $\zeta^0$ meson will be primarily produced at the LHC in gluon-gluon fusion with the largest cross section for the $^1{S_0}$ ground state, and this state will decay into $gg,\gamma\gamma, ZZ^*, WW^*, c\bar c, b\bar b$, $\tau^+\tau^-$ and $\mu^+\mu^-$. At leading order and in the narrow-width approximation, the production cross section for a boson $P$ in $pp$ collisions is given in terms of the gluonic decay width by
\begin{equation}
\label{prodcrosssection}
\sigma(pp\rightarrow P+X)=\frac{\pi^2}{8m_P^3}\Gamma(P\rightarrow gg)\int^1_{\tau_P} dx\frac{\tau_P}{x}g(x,Q^2)g(\tau_P/x,Q^2),
\end{equation}
where $g(x,Q^2)$ is the gluon parton distribution function, and $\tau_P=m_{P}^2/s$ where $s$ is the pp collision energy squared. We identify $P$ with either the pseudoscalar meson $P=\zeta^0$ or the Higgs particle $P=H^0$.

We can obtain approximate results for the $\zeta^0$ boson decay widths using the colored non-relativistic heavy quark model~\cite{Martin,REllis}. The quark wave function for the lowest lying S-wave state is
\begin{equation}
\psi(r)=\frac{1}{\sqrt{4\pi}}R_0(r).
\end{equation}

The interesting two-body decays of $\zeta^0$ will dominate all other annihilation decay channels. We treat $\zeta^0$ as a non-relativistic bound state with velocities $v\rightarrow\, 0$ and with the quarks in an S-wave. The partial width for $\zeta^0\rightarrow A+B$ is given by
\begin{equation}
\Gamma(\zeta^0\rightarrow A+B)=\frac{3\beta}{32\pi^2m_\zeta^2}\vert R_0(0)\vert^2\frac{1}{1+\delta_{AB}}\sigma_{\lambda_A,\lambda_B}
\vert M^{\lambda_A,\lambda_B}(\zeta^0\rightarrow AB)\vert^2_{v=0},
\end{equation}
where
\begin{equation}
\beta=\biggl(\biggl(1-\frac{m_A^2+m_B^2}{m_\zeta^2}\biggr)^2-\frac{4m_A^2m_B^2}{m_{\zeta}^4}\biggr)^{1/2}
\end{equation}
is the standard phase factor and $1/(1+\delta_{AB})$ is a statistical factor.

In the Coulomb approximation, we have
\begin{equation}
\frac{\vert R_0(0)\vert^2}{m_{\zeta^0}}=C_i{\alpha_s}^3m_{\zeta^0},
\end{equation}
where $C_i$ is the quark charge and color factor. For the decay $\zeta^0\rightarrow 2\gamma$, we get to leading order in the Born approximation:
\begin{equation}
\label{pseudogammadecay}
\Gamma(\zeta^0\rightarrow \gamma\gamma)=\biggl(\frac{\alpha}{m_{\zeta^0}}\biggr)^2\vert R_0(0)\vert^2= C_{\gamma\gamma}\alpha^2{\alpha_s}^3m_{\zeta^0}.
\end{equation}

We can compare the leading order partial decay width of $\zeta^0\rightarrow 2\gamma$ to the partial width of the scalar Higgs particle decay in the light mass Higgs limit~\cite{JEllis,Shifman,Marciano}:
\begin{equation}
\Gamma(H^0\rightarrow\gamma\gamma)=\vert I\vert^2\biggl(\frac{\alpha}{4\pi}\biggr)^2\frac{G_Fm_H^3}{8\sqrt{2}\pi},
\end{equation}
where $G_F=1.6637\times 10^{-5}\,{\rm GeV^{-2}}$ is Fermi's constant. In the limit of a Higgs mass $m_H\sim 125$ GeV and for the W and top quark loop contributions, $I_{(W+t)}\sim - 10$, we obtain
\begin{equation}
\label{Higgsgamma}
\Gamma(H^0\rightarrow \gamma\gamma)\sim 31\, {\rm keV}.
\end{equation}

For the gluon final state, we get for the $\zeta^0$ meson decay:
\begin{equation}
\label{gluons}
\Gamma(\zeta^0\rightarrow\, gg)=\frac{4}{3}\biggl(\frac{\alpha_s}{m_{\zeta^0}}\biggr)^2\vert R(0)\vert^2= \frac{4}{3}C_{gg}{\alpha_s}^5m_{\zeta^0}.
\end{equation}
The partial decay width in the Born approximation for the decay $H^0\rightarrow gg$ is given
by
\begin{equation}
\label{Higgsgluons}
\Gamma(H^0\rightarrow gg)=\frac{G_F{\alpha}_s^2m_{H^0}^3}{36\sqrt{2}\pi^3}.
\end{equation}
For $m_{\zeta^0}=m_{H^0}\sim 125$ GeV we obtain the ratio
\begin{equation}
\mu\equiv\frac{\sigma_\zeta}{\sigma_H}=\frac{\Gamma(\zeta^0\rightarrow \gamma\gamma)}{\Gamma(H^0\rightarrow \gamma\gamma)},
\end{equation}
where $\sigma_\zeta$ and $\sigma_H$ denote the production cross sections for a $\zeta^0$ and a Higgs particle, respectively. The value of $\mu$ can be greater than unity. Measurements of the total production cross sections $\sigma_\zeta$ and $\sigma_H$ at the LHC can be used to distinguish between the $\zeta^0$ quark-antiquark pseudoscalar resonance and the standard model Higgs particle.

Hopefully, the 2012 LHC experimental runs will determine whether the hinted excess of events at 125 GeV in the $\gamma\gamma$ decay channel observed by the CMS and ATLAS detectors is accompanied by any excess of events in the $WW^*$, $ZZ^*$, $b\bar b$ and lepton decay channels. If the observed resonance bump at 125 GeV can be verified with sufficient statistical certainty, then an experimental comparison of the branching ratios and production cross sections predicted by a heavy pseudoscalar quarkonium $\zeta^0$ particle and a Higgs particle can resolve whether a $\zeta^0$ resonance or a light Higgs particle has been discovered at the LHC.  If it is found that the 125 GeV resonance is a pseudoscalar quarkonium resonance, then it is explained as a predicted mixing of $\vert\zeta^0\rangle$ and $\vert\zeta^{0'}\rangle$ quark states in QCD. This will mean that the 125 GeV resonance bump cannot be identified with a standard model Higgs particle.

\end{fmffile}

\section*{Acknowledgements}

I thank John Dixon, Bob Holdom, Martin Green, Viktor Toth and Chris Wymant for discussions. This research was generously supported by the John Templeton Foundation. Research at the Perimeter Institute for Theoretical Physics is supported by the Government of Canada through NSERC and by the Province of Ontario through the Ministry of Research and Innovation (MRI).


\begin{thebibliography}{100}


\bibitem{ATLAS1} ATLAS Collaboration, arXiv:1202.1408 [hep-ex].

\bibitem{CMS1} CMS collaboration, arXiv:1202.1488 [hep-ex].

\bibitem{ATLAS2} ATLAS Collaboration, arXiv:1202.1414 [hep-ex].

\bibitem{Tevatron} T. Aaltonen et al. arXiv: 1203.5815 [hep-ex].

\bibitem{pdg} http://pdg.lbl.gov/

\bibitem{Weinberg} S. Weinberg, Phys. Rev. Lett. {\bf 19}, 1267 (1967).

\bibitem{Salam} A. Salam, {\it Elementary Particle Physics} ed. N. Svartholm (Stockholm: Almqvist and Wiksells) 1968.

\bibitem{Moffat} J.W. Moffat, Eur. Phys. J. Plus, 126:53 (2011), arXiv:1006.1859 [hep-ph].

\bibitem{Moffat2} J. W. Moffat, arXiv:1104.5706 [hep-ph].

\bibitem{Moffat3} J. W. Moffat, arXiv:1109.5383 [hep-ph].

\bibitem{Moffat4} J. W. Moffat, arXiv:1203.1573 [hep-ph].

\bibitem{Holdom} B. Holdom, Phys. Lett. {\bf B709}, 381 (2012) [hep-ph].

\bibitem{Gabrielli} E. Gabrielli, B. Mele and M. Raidal, arXiv:1202.1796 [hep-ph]].

\bibitem{Gerard} E. Cerver\'o and J. G\'erard, arXiv:1202.1973 [hep-ph].

\bibitem{Martin} S. P. Martin, Phys. Rev. {\bf D77}, 075002 (2008), arXiv:0801.0237v2 [hep-ph].

\bibitem{REllis} R. K. Ellis, W. J. Stirling and B. R. Webber, {\it QCD and Collider Physics}, Cambridge Monographs on Particle Physics, Nuclear Physics and Cosmology, Cambridge University Press, 1996.

\bibitem{JEllis} J. Ellis, M. K. Gaillard, and D. V. Nanopoulos, Nucl. Phys. {\bf B106}, 292 (1976).

\bibitem{Shifman} M. A. Shifman, A. I. Vainshtein, M. B. Voloshin and V. I. Zakharov, Sov. J. Nucl. Phys. {\bf 30}, 711 (1979) [Yad. Fiz. {\bf 30}. 1368 (1979)].

\bibitem{Marciano} W. J. Marciano, C. Zhang, and S. Willenbrock, arXiv:1109.5304v2 [hep-ph].

\end{thebibliography}
\end{document}